\documentclass[a4paper,11pt,titlepage,nofootinbib]{revtex4-2}
\pdfoutput=1
\usepackage[T1]{fontenc}
\usepackage{natbib}
\usepackage{soul,xcolor}
\usepackage{diagbox}
\usepackage{mathrsfs,bm}
\usepackage{comment}
\usepackage{multirow,booktabs,graphicx}
\usepackage{amsmath,amssymb}
\usepackage[bookmarks=false,
breaklinks=false,pdfborder={0 0 1},colorlinks=false]
{hyperref}
\hypersetup{
colorlinks,linkcolor=blue,citecolor=blue,urlcolor=blue}
\def\d{{\rm d}}

\newcommand{\beq}{\begin{equation}}
\newcommand{\eeq}{\end{equation}}
\newcommand{\bal}{\begin{aligned}}
\newcommand{\eal}{\end{aligned}}
\newcommand{\bin}[2]{\binom{#1}{#2}}
\newcommand{\pochh}[2]{\left(#1\right)_{#2}}

\DeclareMathOperator{\sgn}{sgn}

\begin{document} 

\title{\boldmath On the logarithmic Love number of black holes beyond general relativity}

\author{Sebastian Garcia-Saenz,$^{a}$}
\email{sgarciasaenz@sustech.edu.cn}

\author{Hongbo Lin$^{a}$}
\email{12211826@mail.sustech.edu.cn}

\affiliation{
$^{a}$Department of Physics, Southern University of Science and Technology, \\
Shenzhen 518055, China
}

\begin{abstract}
Tidal Love numbers and other response coefficients of black holes sometimes exhibit a logarithmic dependence on scale, or `running'. We clarify that this coefficient is directly calculable from the structure of the equation obeyed by the field perturbation, and requires no knowledge of the full solution. The derived formula allows us to establish some general results on the existence of logarithmic running. In particular, we show that any static and spherically symmetric spacetime that modifies the Schwarzschild or Reissner-Nordstr\"om solutions in a perturbative way must have non-zero logarithmic Love numbers. This applies for instance to all regular black hole metrics. On the other hand, our analysis highlights the importance of the perturbativity assumption: without it, we find explicit black hole solutions beyond general relativity with exactly zero running. We also illustrate the advantage of our method by recovering and extending the known results for the Hayward metric.
\end{abstract}

\maketitle
\flushbottom

\section{Introduction}
\label{sec:intro}

Tidal Love numbers characterize the deformation of an object under an external gravitational field~\cite{Hinderer:2007mb,Damour:2009vw,Binnington:2009bb}. They arise as the linear response coefficients of the gravitational potential or, more accurately, the spacetime metric. As such, they are quantities that encode information both of the object and the gravitational theory, providing an interesting observational target in the astrophysical context~\cite{Flanagan:2007ix,Cardoso:2017cfl,ET:2019dnz,Piovano:2022ojl}. Of special interest is the question on the Love numbers of black holes. A celebrated result, one which has attracted much interest in recent work, is the property that black holes in general relativity (GR) have exactly zero Love numbers~\cite{Fang:2005qq,Hui:2020xxx,LeTiec:2020bos,LeTiec:2020spy,Chia:2020yla,Poisson:2021yau,Charalambous:2021mea}.\footnote{For exceptions, see e.g.\ \cite{Emparan:2017qxd,Garcia-Saenz:2022wsl,Nair:2024mya,Chakraborty:2025zyb,Pang:2025myy}.} This fact is all the more intriguing when one notices that Love numbers appear as Wilson coefficients in the effective field theory (EFT) of point particles~\cite{Goldberger:2004jt,Goldberger:2007hy,Rothstein:2014sra,Porto:2016pyg}, so that their vanishing appears at first sight as a fine-tuning problem~\cite{Porto:2016zng}. The remarkable resolution is provided by some hidden symmetries of the equations for perturbations~\cite{Penna:2018gfx,Charalambous:2021kcz,Hui:2021vcv,Hui:2022vbh,BenAchour:2022uqo,Charalambous:2022rre,Combaluzier-Szteinsznaider:2024sgb,Rai:2024lho,DeLuca:2024nih,Berens:2025okm,Parra-Martinez:2025bcu,DeLuca:2025zqr}, although the full implications of these are yet to be understood.

The flip-side to this `no-Love' theorem is that any violation of it would unambiguously indicate a deviation from GR. It is therefore pertinent to ask precisely how such deviations would imprint themselves in the Love numbers. In this paper we make progress on this question by focusing on a particular aspect of Love numbers, namely their logarithmic part. Indeed, it is an interesting fact that Love numbers are not necessarily purely constants but may include a logarithmic dependence on scale. Such `running' of Love numbers is of course reminiscent of the consequences of renormalization in quantum field theory, and indeed it has been shown that in the point-particle EFT the Love numbers are generically divergent and must be renormalized~\cite{Kol:2011vg,Ivanov:2022hlo,Saketh:2023bul,Mandal:2023hqa,Hadad:2024lsf,Kobayashi:2025vgl,Caron-Huot:2025tlq,Kosmopoulos:2025rfj}. What is remarkable is that this occurs already at tree level. This highlights the importance of the logarithmic term: it is a beta function, and as such it is an unambiguous and well-defined observable, unlike the constant terms~\cite{Barbosa:2025uau}.

A perhaps underappreciated fact is that the logarithmic static Love number is directly calculable from the Fuchsian theory of linear differential equations. In other words, one does not need to know the full solution of the equation, nor a particular solution consistent with some boundary condition. Even in cases when such solutions are available, the resulting formula bypasses the need to deal with special functions and related connection identities. The method is also inherently non-perturbative, in the sense that it does not rely on a `seed' equation with known solution, which is the most common approach to calculate Love numbers of black holes beyond GR~\cite{Barura:2024uog,Kobayashi:2025swn,Coviello:2025pla,Cano:2025zyk,Wang:2025oek,Liu:2025iby}. Thanks to the explicitness and universality of the derived formula, we are able to establish some general results on the logarithmic Love numbers of general black hole metrics. We show for instance that any static and spherically symmetric spacetime that modifies the Schwarzschild or Reissner-Nordstr\"om (RN) solutions in a perturbative way must have non-zero running beyond some multipole order, whose sign can be generally predicted. Black hole metrics that are not perturbatively close to Schwarzschild or RN may however evade this outcome, and we provide an explicit example.

Our analysis, we clarify, is restricted to the calculation of static Love numbers in the context of static, spherically symmetric, asymptotically flat spacetimes in four dimensions (with the exception of the example of the Schwarzschild-Tangherlini metric). We consider two classes of equations: the Klein-Gordon equation for a probe scalar field and the (modified) Regge-Wheeler equation for odd-parity tensor perturbations. These assumptions are made for the sake for simplicity, since our main aim is to provide a proof of concept rather than concrete results, although we do illustrate the advantages of the approach with some explicit calculations. Our methods are however applicable much more broadly, and we offer some comments on potential extensions in the final discussion section.


\section{Logarithmic Love number}

\subsection{General formula for the logarithmic Love number}

Consider a general, homogeneous, linear second-order differential equation,
\beq
\hat{L}u:=u''(x)+\frac{p(x)}{x}u'(x)+\frac{q(x)}{x^2}u(x)=0 \,.
\eeq
Suppose the functions $p(x)$ and $q(x)$ are analytic at $x=0$, i.e.\ $x=0$ is a regular singular point of the equation. From Fuchsian theory (see e.g.~\cite{Bender:1999box,boyce2004ede}), we know that typically the local solution near $x=0$ will be of the form of a Frobenius series with leading powers, $r_1$ and $r_2$, as determined by the indicial equation:
\beq
R(r):=r^2+(p(0)-1)r+q(0)=0 \,.
\eeq

The Frobenius series fails to produce two linearly independent solutions in the case when $r_1-r_2=N$, a positive integer (we assume $r_1>r_2$). In this situation, we construct the first solution by assuming a Frobenius series:
\beq
u_1(x):=x^{r_1}\sum_{n=0}^{\infty}a_nx^n \,.
\eeq
The sequence $\{a_n\}$ is determined by the recurrence relation
\beq
R(r_1+n)a_n+\sum_{m=0}^{n-1}\left[(r_1+m)p_{n-m}+q_{n-m}\right]a_m=0 \,,\qquad n\geq1 \,,
\eeq
where $p_n$ and $q_n$ denote the $n$-th Taylor coefficients of $p(x)$ and $q(x)$, respectively. Note that $a_0$ is undetermined at this stage.

To construct a second solution, consider the series
\beq
u_r(x):=x^{r}\sum_{n=0}^{\infty}a_n(r)x^n \,,
\eeq
where the sequence $\{a_n(r)\}$ is determined by the recurrence relation
\beq
R(r+n)a_n(r)+\sum_{m=0}^{n-1}\left[(r+m)p_{n-m}+q_{n-m}\right]a_m(r)=0 \,,\qquad n\geq1 \,,
\eeq
in terms of $a_0(r)\equiv a_0(r_1)\equiv a_0$ and the parameter $r$. Note that here $r$ is left arbitrary, so that $u_r(x)$ is not a solution of the differential equation except for $r=r_{1,2}$. The failure of $u_r(x)$ to satisfy the equation may be expressed as
\beq
\hat{L}u_r(x)=a_0x^{r-2}R(r) \,,
\eeq
and recall that, by definition, $R(r_{1,2})=0$.

Consider next the function $\frac{\partial u_r}{\partial r}\big|_{r=r_1}$, given explicitly by
\beq
\frac{\partial u_r}{\partial r}\bigg|_{r=r_1}=u_1(x)\log x+x^{r_1}\sum_{n=0}^{\infty}\frac{\partial a_n(r)}{\partial r}\bigg|_{r=r_1}x^n \,,
\eeq
and apply the operator $\hat{L}$ to it. The result is
\beq \label{eq:frob1}
\hat{L}\frac{\partial u_r}{\partial r}\bigg|_{r=r_1}=a_0R'(r_1)x^{r_2+N-2} \,,
\eeq
recalling that $r_1=r_2+N$. Note that at this step we assume $r_1\neq r_2$ (otherwise $R'(r_1)=0$). We can regard \eqref{eq:frob1} as an inhomogeneous equation, whose homogeneous part corresponds to the equation we are trying to solve. Eq.\ \eqref{eq:frob1} shows, obviously, that $\frac{\partial u_r}{\partial r}\big|_{r=r_1}$ is a particular solution. A second particular solution is given by
\beq
\widetilde{u}_2(x):=x^{r_2}\sum_{n=0}^{\infty}b_nx^n \,,
\eeq
where the sequence $\{b_n\}$ is determined by the recurrence relation
\beq \label{eq:frob3}
R(r_2+n)b_n+\sum_{m=0}^{n-1}\left[(r_2+m)p_{n-m}+q_{n-m}\right]b_m=0 \,,\qquad n\geq1 \,,\qquad n\neq N\,,
\eeq
and
\beq \label{eq:frob2}
\sum_{m=0}^{N-1}\left[(r_2+m)p_{N-m}+q_{N-m}\right]b_m=a_0R'(r_1) \,.
\eeq
The last relation determines $a_0$ in terms of the $\{b_n\}$. The two coefficients $b_0$ and $b_N$ are undetermined and correspond to the integration constants of the general solution.

Neither $\frac{\partial u_r}{\partial r}\big|_{r=r_1}$ nor $\widetilde{u}_2$ are solutions of the original equation. But since they both solve the same inhomogeneous equation, their difference is the correct solution we are looking for:
\beq
u_2(x):=u_1(x)\log x+x^{r_1}\sum_{n=0}^{\infty}\frac{\partial a_n(r)}{\partial r}\bigg|_{r=r_1}x^n-x^{r_2}\sum_{n=0}^{\infty}b_nx^n \,.
\eeq

Eq.\ \eqref{eq:frob2} is an important relation. It gives us an explicit formula for the leading coefficient of the logarithm in the general solution:\footnote{It is worth emphasizing that $a_0$ is completely unambiguous modulo trivial rescalings (which we will fix later by choosing $b_0=1$). In particular, $a_0$ will not change when performing analytic changes of variable of the form $x\mapsto x+cx^2+\cdots$, i.e.\ the type of redefinitions which lead to ambiguities in the definition of the Love number when the logarithm is absent (see e.g.~\cite{Hui:2020xxx,Pereniguez:2021xcj}).}
\beq
a_0=\frac{1}{R'(r_1)}\sum_{m=0}^{r_1-r_2-1}\left[(r_2+m)p_{r_1-r_2-m}+q_{r_1-r_2-m}\right]b_m \,.
\eeq
Of course, this is still not a closed-form result, because one must first determine the constants $\{b_1,b_2,\ldots ,b_{r_1-r_2-1}\}$ using the recurrence relation \eqref{eq:frob3}. The latter does not seem to admit a closed form solution given general $\{p_n\}$ and $\{q_n\}$, however it can be explicitly solved in some particular cases of physical interest.

\subsection{Klein-Gordon equation on a black hole background}

The simplest system in which one can investigate the linear response of a black hole is given by a probe scalar field. The relevant equation is then the Klein-Gordon equation on the background of a general static and spherically symmetric metric,
\beq \label{eq:general metric fg}
\d s^2=-f(r)\d t^2+\frac{\d r^2}{g(r)}+r^2\d\Omega_2^2 \,.
\eeq
It is useful to think of this line element as a deformation of the Schwarzschild metric, although later we will derive some non-perturbative results. To this end we parameterize the functions $f$ and $g$ as power series of the following form:
\beq
f(x)=1-x\left(1+\sum_{n=1}^{\infty}f_nx^n\right) \,,\qquad g(x)=1-x\left(1+\sum_{n=1}^{\infty}g_nx^n\right) \,,
\eeq
where $x:=r_g/r$ and $r_g$ is the Schwarzschild radius. The Schwarzschild metric then corresponds to the particular case where $f_n=g_n=0$ for all $n$, while the Reissner-Nordstr\"om metric is given by the choice $f_1=g_1\neq0$, $f_n=g_n=0$ $\forall n>1$.

We consider a static (i.e.\ time-independent) scalar perturbation and separate the massless Klein-Gordon equation, $\Box\Phi=0$, in spherical harmonics,
\beq
\Phi=x\sum_{l,m}u(x)Y^{l,m}(\Omega) \,,
\eeq
and we will omit the $l,m$ labels on the mode function $u$. One finds
\beq \label{eq:KG u}
u''+\frac{1}{2}\left(\frac{4}{x}+\frac{f'}{f}+\frac{g'}{g}\right)u'+\frac{1}{2x}\left(\frac{f'}{f}+\frac{g'}{g}-\frac{2l(l+1)}{xg}\right)u=0 \,,
\eeq
where primes denote differentiation with respect to $x$.

It is easy to check that $x=0$ is a regular singular point of Eq.\ \eqref{eq:KG u}. In the notations of the previous subsection, we have $r_1=l$ and $r_2=-(l+1)$, i.e.\ $r_1-r_2=2l+1$, so we are in the situation explained above and we generically expect the presence of a logarithmic contribution to the Love number, i.e.\ a non-zero $a_0$. Notice that $R'(r_1)=2l+1$ for this equation. More precisely, we define the logarithmic Love number as the ratio $a_0/b_0$. This ratio is, of course, independent of the integration constant $b_0$, so it is a characteristic property of the system. Without loss of generality we simply choose $b_0=1$ in what follows. Eq.\ \eqref{eq:frob2} then reduces to
\beq \label{eq:a0 formula}
a_0(l)=\frac{1}{2l+1}\sum_{m=0}^{2l}\left[(m-l-1)p_{2l+1-m}+q_{2l+1-m}\right]b_m \,,
\eeq
with the coefficients $\{b_1,b_2,\ldots ,b_{2l}\}$ determined recursively by the recurrence relation
\beq \label{eq:b recurrence}
n(n-2l-1)b_n+\sum_{m=0}^{n-1}\left[(m-l-1)p_{n-m}+q_{n-m}\right]b_m=0 \,.
\eeq
Note that the $\{b_n\}$ and $\{q_n\}$ are also implicitly functions of $l$.

\subsection{Schwarzschild-Tangherlini black hole}

Although not within the main scope of this paper, the linear response of a Schwarzschild-Tangherlini black hole (i.e.\ the extension of the Schwarzschild metric to dimension $d>4$) provides a clean and well-understood setting for testing our result. In particular, it is known that perturbations with multipole number $l$, such that $l/(d-3)$ is a half-integer, are characterized by a non-zero logarithmic Love number~\cite{Kol:2011vg,Cardoso:2019vof}. In Ref.~\cite{Hui:2020xxx} this was proved through an analysis of the general solution expressed in a suitable basis of hypergeometric functions, and the identification of the solution which is non-singular at the event horizon. Our method allows one to bypass this procedure, relying on an explicit algebraic formula that requires no knowledge of the general solution or the singularity structure thereof.

Scalar perturbations on a Schwarzschild-Tangherlini background are governed by the equation~\cite{Hui:2020xxx}
\beq
u''+\frac{1}{2}\left(\frac{4}{x}+\frac{2f'}{f}\right)u'+\frac{1}{2x}\left(\frac{(d-2)f'}{f}-\frac{l(l+d-3)}{xf}-\frac{(d-4)(d-2)}{2}\right)u=0 \,,
\eeq
which indeed reduces to \eqref{eq:KG u} when $d=4$ and $g=f$. Here we have $f=1-x^{d-3}$. From the equation we easily infer
\begin{equation*}
	p_n=\begin{cases}
		2 \,,   & \text{if $n=0$} \,,         \\
		3-d \,, & \text{if $(d-3)\mid n$} \,,   \\
		0 \,,  & \text{otherwise} \,,
	\end{cases} \qquad\qquad
	q_n=\begin{cases}
		-l(l+d-3)-\frac{(d-4)(d-2)}{4} \,, & \text{if $n=0$} \,,             \\
		-\frac{(d-3)(d-2)}{2}-l(l+d-3) \,, & \text{if $(d-3)\mid n$} \,,     \\
		0 \,,                             & \text{otherwise} \,,
	\end{cases}
\end{equation*}
where $m\mid n$ means that $m$ divides $n$. The indicial equation is also modified: $R(r)=r^2+r-l(l+d-3)-\frac{(d-4)(d-2)}{4}$, with roots $r_1=-2+\frac{d}{2}+l$ and $r_2=1-\frac{d}{2}-l$, i.e.\ $r_1-r_2=2l+d-3=R'(r_1)$.

Eq.\ \eqref{eq:b recurrence} for $b_n$ may be simplified to
\begin{equation}
	n(n-2l-d+3)b_n=\sum_{m=0,(d-3)\mid(n-m)}^{n-1}\left[(d-3)m+l^2\right]b_m \,.
\end{equation}
The condition $(d-3)\mid(n-m)$ translates into $m=n-(d-3)k$ ($k\in\mathbb{N}^+$), implying that $b_1=b_2=\cdots=b_{d-4}=0$. Furthermore, in the above recurrence $b_n$ only depends on the $b_m$ such that $n= m\,\text{mod}\,(d-3)$. Thus $b_{(d-3)k}\neq 0$ ($k\in\mathbb{N}$), otherwise $b_n=0$. This allows us to recast the equation as
\begin{equation}
	n(d-3)\left[(n-1)(n-3)-2l\right]b_{(d-3)n}=\sum_{k=0}^{n-1}\left[(d-3)^2k+l^2\right]b_{(d-3)k} \,.
\end{equation}
The solution may now be found through standard methods:
\begin{equation}
	b_{(d-3)n}=\prod_{k=0}^{n-1}\frac{\left[(d-3)k-l\right]^2}{(k+1)(d-3)\left[(d-3)k-2l\right]}=\frac{\left(-\frac{l}{d-3}\right)^2_n}{n!\left(-\frac{2l}{d-3}\right)_n} \,,
\end{equation}
where $(x)_n:=\prod_{k=0}^{n-1}(x+k)$ is the rising factorial or Pochhammer symbol.\footnote{Notice that $(x)_n$ is defined for any real $x$. The subscript argument $n$ will always be assumed to be a non-negative integer, and we define $(x)_0=1$.}

Using a similar reasoning we may next express the formula for the logarithmic Love number, Eq.\ \eqref{eq:a0 formula}, as
\begin{equation}
	a_0=\frac{1}{2l+d-3}\sum_{m=0,(d-3)\mid(2l+d-3-m)}^{2l+d-4}\left[(d-3)m+l^2\right]b_m \,.
\end{equation}
The terms entering in the sum have $m=2l+(1-k)(d-3)$ ($k\in\mathbb{N}^+$), while for non-zero $b_m$ we require $m=(d-3)j$ ($j\in\mathbb{N}$). These two conditions imply $(d-3)\mid 2l$, otherwise $a_0=0$. This shows that $l/(d-3)$ must be an integer or half-integer. In that case, letting $2l=(d-3)L$ ($L\in\mathbb{N}^+$), we find
\begin{equation}
a_0=\frac{l^2}{(L+1)(d-3)}b_{(d-3)L}=\frac{(-1)^L(d-3)}{L!(L+1)!}\left(-\frac{L}{2}\right)^2_{L+1} \,.
\end{equation}
If $\frac{L}{2}=\frac{l}{d-3}$ is an integer then $\left(-\frac{L}{2}\right)_{L+1}=0$. Otherwise $\left(-\frac{L}{2}\right)_{L+1}\neq0$ and therefore $a_0\neq0$. This demonstrates, as claimed, that $l/(d-3)$ must be a half-integer. Furthermore, we have $\left(-\frac{L}{2}\right)^2_{L+1}=\frac{\Gamma(\frac{l}{d-3}+1)^2}{\Gamma(-\frac{l}{d-3})^2}$, and we find the derived $a_0$ to be in precise agreement with the results of~\cite{Kol:2011vg,Hui:2020xxx}.

\subsection{Tensor perturbations}

The extension of the above analysis to other types of probe fields (e.g.\ spin-1 gauge field) is straightforward. Of most interest is however the linear response of metric perturbations, corresponding to the tidal Love numbers of the system. For black holes beyond GR, or for non-vacuum black holes, the probe limit is in general inconsistent since neglecting the backreaction of perturbations would be equivalent to neglecting the source term in the Einstein equation. The equations for tensor perturbations are therefore determined by the theory and do not have a universal structure.

To address the question in a model-independent way, we resort to the effective field theory (EFT) for black hole perturbations with time-like scalar profile in scalar-tensor theory~\cite{Mukohyama:2022enj,Mukohyama:2022skk,Barura:2024uog} (see also \cite{Kobayashi:2012kh,Franciolini:2018uyq,Kuntz:2019zef,Hui:2021cpm,Mukohyama:2023xyf}). This description covers a wide class of theories of modified gravity, including ones that accommodate regular black hole solutions. We restrict our attention to the odd-parity sector, described by a single dynamical mode which satisfies a modified Regge-Wheeler equation.\footnote{To our knowledge, the even-parity sector equations have not been worked out in this context. It will presumably include additional model-dependent functions. The reason for us to focus on the odd-parity sector is that the resulting mode equation is known explicitly, and is sufficiently simple and generic for our present scope.} In the static limit the equation is given by~\cite{Mukohyama:2022skk}
\beq \label{eq:modified RW}
\left[x^2F(x)\frac{\d}{\d x}\left(x^2F(x)\frac{\d}{\d x}\right)-F(x)\tilde{V}(x)\right]u(x)=0 \,,
\eeq
with the definitions
\beq\bal
F&:= \sqrt{\frac{g}{f}}\frac{f+\alpha_T}{\sqrt{1+\alpha_T}} \,,\\
\tilde{V}&:= \sqrt{\frac{f}{g}}\sqrt{1+\alpha_T}\left[l(l+1)-2\right]x^2+\frac{x}{(1+\alpha_T)^{1/4}}\frac{\d}{\d x}\left[x^2F(x)\frac{\d}{\d x}\left(x(1+\alpha_T)^{1/4}\right)\right] \,.
\eal\eeq
Here $\alpha_T(x)$ is a model-dependent function representing the deviation of the speed of gravitational waves from unity, thus serving as a sensitive observational target. For instance, $\alpha_T=0$ for the Schwarzschild metric and $\alpha_T=-\frac{\eta x^3(\eta x^3+2)}{(\eta x^3+1)^2}$ for the Hayward metric, where $\eta$ is a constant~\cite{Barura:2024uog}. By assumption, $\alpha_T\to0$ as $x\to0^+$ in order to recover the GR behavior at infinity. We will further assume $\alpha_T$ is analytic at $x=0$. Under these conditions, we find that the indicial equation for \eqref{eq:modified RW} is the same as in the probe field case. Therefore the master formula \eqref{eq:a0 formula} applies also in the tensor case, although of course $p_n$ and $q_n$ will have a different functional dependence on the background metric and will further also depend on the function $\alpha_T$.

\section{Probe scalar field}

In this section we study the application of the master formula for the running of the Love number, Eq.\ \eqref{eq:a0 formula}, in the case of a scalar field in the probe approximation, Eq.\ \eqref{eq:KG u}.

\subsection{Monopolar response}

It is instructive to consider first the simplest case of monopole perturbations, i.e.\ $l=0$. Eq.\ \eqref{eq:a0 formula} then reduces to
\beq
a_0(0)=q_1-p_1 \,,
\eeq
and a direct calculation produces $p_1=q_1=-1$, independent of the coefficients $f_n$ and $g_n$. Thus, for any metric, the logarithmic Love number vanishes for monopole modes.

This result is actually an immediate corollary of a more general statement: the full Love number vanishes for $l=0$. Indeed, in this particular case Eq.\ \eqref{eq:KG u} reads
\beq
u''+\left(\frac{2}{x}+\frac{(fg)'}{2fg}\right)u'+\frac{(fg)'}{2xfg}u=0 \,,
\eeq
which admits the exact solution
\beq
u=\frac{v_0}{x}+\frac{w_0}{x}\int^x_{x_h}\frac{\d \bar{x}}{\sqrt{fg}} \,,
\eeq
with integration constants $v_0$ and $w_0$, which we have chosen so that the lower limit of the integral, $x_h$, corresponds to the event horizon of the black hole. By assumption, the metric functions $f$ and $g$ vanish at the horizon, $f(x)\simeq f'(x_h)(x-x_h)$ and $g(x)\simeq g'(x_h)(x-x_h)$ (we assume for simplicity a non-extremal black hole, but the extremal case yields the same conclusion). Thus, in the vicinity of the horizon, $x=x_h+\delta x$, we have
\beq
u= \frac{v_0}{x}+\frac{w_0}{x}\int^{x_h+\delta x}_{x_h}\frac{\d\bar{x}}{\sqrt{fg}}\simeq \frac{v_0}{x}+\frac{w_0}{x}\frac{1}{\sqrt{f'(x_h)g'(x_h)}}\int^{\delta x}_{0}\frac{\d \tilde{x}}{\tilde{x}} \,,
\eeq
which clearly diverges unless $w_0=0$. The resulting regular solution, $u\propto 1/x$, is purely growing at infinity and thus corresponds to a source term with vanishing response, i.e.\ a zero Love number.

\subsection{Perturbative deformation of the Schwarzschild metric} \label{subsec:scalar deformation}

\subsubsection{Single-term deformation with $f=g$}

We begin with a study of line elements with $f=g$ and assume the metric to be perturbatively close to the Schwarzschild metric everywhere outside the event horizon, in the sense that
\beq \label{eq:single term metric fg}
f=g=1-x\left(1+\alpha x^N\right) \,,
\eeq
where $N$ is a positive integer and $\alpha$ is a small parameter, such that $|\alpha|x^N\ll 1$ everywhere in the domain of interest. The metric defined by \eqref{eq:single term metric fg} should be interpreted as a truncation of a generically infinite Taylor series,\footnote{Of course this series will in general have a finite radius of convergence. For the Frobenius method to be applicable everywhere in the black hole exterior, we must therefore assume that $p$ and $q$ admit Taylor series expansions about $x=0$ with a domain of convergence that contains the event horizon. Notice that this will be true if $p$ and $q$ are real-analytic in a (connected) domain that contains $x=0$ and the horizon, so generically valid for regular enough metrics.} so that our results will be valid to $\mathcal{O}(\alpha)$. Nevertheless, we will also prove some results at arbitrarily high order in $\alpha$,\footnote{This would be relevant for metrics such that the Taylor series of $f$ skips some powers of $\alpha$. For instance, if we have $f=1-x\left(1+\alpha x^N+\alpha^3 x^{3N}+\cdots\right)$, then it would be consistent to calculate results at $\mathcal{O}(\alpha^2)$ with the truncation \eqref{eq:single term metric fg}. We admit however that we are not aware of any physical example of this kind.} for the simple reason that it is relatively straightforward to do so.

This set-up applies to a large class of black holes in modified gravity, including most regular black hole metrics routinely studied in the literature (see~\cite{Lan:2023cvz} for a review). Such metrics are precisely constructed to introduce qualitative modifications in the black hole interior, while the outer horizon and exterior domain remain perturbatively close to Schwarzschild. The order $N$ at which the deformation starts characterizes different types of regular black holes. For instance, the Simpson-Visser~\cite{Simpson:2019mud}, Bardeen~\cite{Bardeen:1968a} and Hayward~\cite{Hayward:2005gi} metrics have $N=1,2,3$, respectively. In fact, it is not hard to construct examples with arbitrarily large $N$~\cite{Zhou:2022yio}. Eq.\ \eqref{eq:single term metric fg} also naturally encompasses black hole solutions in the EFT of gravity~\cite{Cardoso:2018ptl,DeLuca:2022tkm,Katagiri:2023umb,Barbosa:2025uau}, where corrections must be small by consistency.\footnote{In the EFT context, one would not expect the series expansions of $f$ and $g$ to have a finite radius of convergence, i.e.\ they correspond to asymptotic series. This does not invalidate the Frobenius method provided we interpret the result for the Love number as an asymptotic series in the coupling constants, as is usual in quantum field theory.} Corrections to vacuum GR start at $N=5$ from curvature cubed terms, or $N=7$ if the former happen to vanish, as they do in some string theories~\cite{Gross:1986iv,Kikuchi:1986rk,Gross:1986mw}.

The simple form \eqref{eq:single term metric fg} of the metric functions allows us to derive closed-form expressions for the coefficients $p_n$ and $q_n$ of Eq.\ \eqref{eq:KG u}. Noticing that
\beq \label{eq:pq when g equals f}
p=2+x\frac{\d}{\d x}\log f \,,\qquad\qquad q=x\frac{\d}{\d x}\log f-\frac{l(l+1)}{f} \,,
\eeq
valid when $g=f$, it is easy to obtain\footnote{The binomial coefficient $\bin{N}{M}\equiv \frac{N!}{M!(N-M)!}$ is defined as zero if $N<0$ or $M>N$.}
\begin{equation*}
p_n^{(r)}=\begin{cases}
2  & \text{if $n = 0$}\,, \\
-\frac{n}{n-rN}\bin{n-rN}{r} & \text{if $n \geq 1$}\,, \\
	\end{cases}\qquad\qquad q_n^{(r)}=\begin{cases}
	-l(l+1) & \text{if $n = 0$}\,, \\
	-\frac{n}{n-rN}\bin{n-rN}{r}-l(l+1)\bin{n-rN}{r} & \text{if $n \geq 1$} \,,
	\end{cases}
\end{equation*}
where $p_n^{(r)}$ is the coefficient of $\alpha^r x^n$ in the double series expansion of $p$ in powers of $\alpha$ and $x$, and similarly for $q$. Substituting into \eqref{eq:b recurrence} yields the recurrence relation
\begin{equation} \label{eqn:scalar-single-full-bn}
n(n-2l-1)b_n-\sum_{m=0}^{n-1}\sum_{r=0}^{\lfloor\frac{n-m}{N}\rfloor}\left[\frac{(m-l)(n-m)}{n-m-rN}+l(l+1)\right]\bin{n-m-rN}{r}\alpha^r b_m=0 \,.
\end{equation}

\noindent\textit{Leading order.---} We focus first on Eq.\ \eqref{eqn:scalar-single-full-bn} at first order in $\alpha$. After a simplification we have
\begin{equation} \label{eqn:scalar-single-order-1-bn}
n(n-2l-1)b_n=\sum_{m=0}^{n-1}(m+l^2)b_m+\alpha\sum_{m=0}^{n-N-1}\left[(m-l)(n-m)+l(l+1)(n-m-N)\right] b_m \,.
\end{equation}
Expanding $b_n=b_n^{(0)}+b_n^{(1)}\alpha+\cdots$ and matching powers of $\alpha$ we arrive at
\begin{equation} \label{eqn:scalar-single-order-1-bn-separated}
	\begin{cases}
		n(n-2l-1)b_n^{(0)}=\displaystyle\sum_{m=0}^{n-1}(m+l^2)b_m^{(0)} \,, \\
		n(n-2l-1)b_n^{(1)}=\displaystyle\sum_{m=0}^{n-1}(m+l^2)b_m^{(1)}+\sum_{m=0}^{n-N-1}\left[(m-l)(n-m)+l(l+1)(n-m-N)\right]b_m^{(0)} \,.
	\end{cases}
\end{equation}
The first recurrence may be solved through standard methods, with the result (recall that we set $b_0=1$)
\begin{equation}
	b_n^{(0)}=\frac{(-1)^n}{(2l)!}\frac{(2l-n)!\pochh{l-n+1}{n}^2}{n!} \,.
\end{equation}
The second recurrence in \eqref{eqn:scalar-single-order-1-bn-separated} is identical except for the inhomogeneous term. Since it is linear, it may be solved via the Green `function' (here a sequence) method. To this end it is useful to first simplify the recurrence, which is achieved by considering the same equation with $n\mapsto n+1$ and subtracting. Eventually we find
\begin{equation} \label{eqn:bn-order-1-recurrence}
	(n+1)(n-2l)b_{n+1}^{(1)}-(n-l)^2b_n^{(1)}=(n-N-l)(n-l)b_{n-N}^{(0)} \,,
\end{equation}
to be solved with the initial condition $b_0^{(1)}=0$. The solution is
\begin{equation}
b_n^{(1)}=\sum_{m=0}^{n-1} g_{m,n} (m-N-l)(m-l)b_{m-N}^{(0)} \,,
\end{equation}
in terms of the Green `function' $g_{m,n}$, defined by the recurrence
\begin{equation}
(n+1)(n-2l)g_{m,n+1}-(n-l)^2g_{m,n}=\delta_{mn} \,,
\end{equation}
with $g_{m,0}=0$. The solution is\footnote{Notice that apparently singular terms, e.g.\ here $1/\pochh{1-l}{m}$, should of course be understood to be evaluated only after expanding the factorials and cancelling common terms in numerator and denominator.}
\begin{equation}
g_{m,n}=\frac{m!}{(m-2l)n!}\frac{(1-2l)_m(1-l)_{n-1}^2}{(1-2l)_{n-1}(1-l)_m^2} \,,
\end{equation}
valid for $n\in[\![m+1,2l]\!]$. Collecting results and simplifying we arrive at
\begin{equation} \label{eqn:bn-order-1-sol}
b_n^{(1)}=\frac{(-1)^{n-N+1}}{(2l)!}\sum_{m=N}^{n-1}\frac{(2l-m+N)!}{(m-N)!}\frac{\pochh{-l}{m-N}\pochh{-l}{m-N+1}\pochh{l+1-n}{n-m-1}\pochh{l+1-n}{n-m}}{\pochh{m+1}{n-m}\pochh{2l+1-n}{n-m}} \,.
\end{equation}
Notice that $b_n^{(1)}=0$ for $n\leq N$ since $b^{(0)}_{n-N}$ in \eqref{eqn:bn-order-1-recurrence} only `sources' $b^{(1)}$ when $n\geq N$.

Continuing with the calculation of the logarithmic Love number, we substitute $b_m$ into the master formula \eqref{eq:a0 formula} after expanding $a_0=a_0^{(0)}+\alpha a_0^{(1)}+\cdots$, explicitly
\begin{equation}
	\begin{cases}
		(2l+1)a_0^{(0)}=\displaystyle -\sum_{m=0}^{2l}(m+l^2)b_m^{(0)} \,, \\
		(2l+1)a_0^{(1)}=\displaystyle -\sum_{m=0}^{2l}(m+l^2)b_m^{(1)}-\sum_{m=0}^{2l-N}[(m-l)(2l+1-m)+(2l+1-m-N)l(l+1)]b_m^{(0)} \,.
	\end{cases}
\end{equation}
A direct computation of the sum in the first equation yields $a_0^{(0)}=0$, i.e.\ a vanishing Love number in the Schwarzschild case, providing a non-trivial check of our calculations. Using the fact that $b_n^{(1)}=0$ for $n\leq2l$ and the properties of $b_n^{(0)}$, the second equation simplifies to
\beq
a_0^{(1)}=-\frac{\alpha}{2l+1}\left[l^2b_{2l}^{(1)}+l(l-N)b_{2l-N}^{(0)}\right] \,,
\eeq
or, more explicitly,
\beq\bal \label{eq:final a0 order alpha}
a_0&=\alpha\frac{(-1)^N}{(2l+1)!}\sum_{m=N}^{2l}\frac{(2l-m+N)!}{(2l-m)!(m-N)!}\frac{\pochh{-l}{m-N}\pochh{-l}{m-N+1}\pochh{-l}{2l-m}\pochh{-l}{2l-m+1}}{\pochh{m+1}{2l-m}}+\mathcal{O}(\alpha^2) \\
&=\alpha\frac{(-1)^N}{(2l+1)!}\sum_{m=0}^{2l-N}\frac{(m-l)(l-m-N)(2l-m)!}{(2l-m-N)!m!}\frac{\pochh{-l}{m}^2\pochh{-l}{2l-m-N }^2}{\pochh{m+N+1}{2l-m-N}}+\mathcal{O}(\alpha^2) \,,
\eal\eeq
where the second line follows after some simple manipulations.

\noindent\textit{Vanishing logarithmic Love.---} With the above closed-form expression at hand, we may ask about necessary and sufficient conditions for $a_0$ to vanish for given values of $l$ and $N$. First, the factorial in the denominator immediately implies that $a_0$ vanishes when $2l<N$ (which has the trivial corollary, already mentioned, that $a_0$ necessarily vanishes for $l=0$). Second, notice that the Pochhammer symbols $\pochh{-l}{m}$ and $\pochh{-l}{2l-m-N}$ vanish when $m\geq l+1$ and $m\leq l-N-1$, respectively, so their product vanishes in the range $l+1\leq m\leq l-N-1$, while the product $(m-l)(l-m-N)$ is non-negative in the complement of this range. Therefore, $a_0$ is non-zero if and only if $[\![l-N+1,l-1]\!]$ is non-empty. These simple observations demonstrate the following theorem.

\noindent\textbf{Theorem 1.} [\textit{Vanishing $a_0$ for $f=g=1-x(1+\alpha x^N+\cdots)$ at $\mathcal{O}(\alpha)$}] $a_0$ vanishes at $\mathcal{O}(\alpha)$ if and only if either $N=1$ or $l\leq \lfloor\frac{N-1}{2}\rfloor$. Otherwise $a_0$ is proportional to $\alpha$ by a non-zero rational number with sign $(-1)^N$.

The conclusion that $a_0=0$ for $N=1$ is consistent with the known result stating that the RN black hole has zero Love number~\cite{Pereniguez:2021xcj,Rai:2024lho,Xia:2025zfp}. As a simple corollary, we see that the logarithmic running of the dipolar ($l=1$) Love number vanishes for $N\geq 3$, which includes the Hayward metric as well as EFT-corrected black holes.

\noindent\textit{Higher order.---} While it is technically challenging to derive closed-form results at arbitrarily high orders in $\alpha$, it is straightforward to infer some general results on the vanishing of $a_0$ at any given $\mathcal{O}(\alpha^r)$.

We denote the $\mathcal{O}(\alpha^r)$ coefficient of $b_m$ in the sum of \eqref{eqn:scalar-single-full-bn} as
\begin{equation}
	B^{(r)}_{n,m,l,N}:=\left[\frac{(m-l)(n-m)}{n-m-rN}+l(l+1)\right]\bin{n-m-rN}{r} \,,
\end{equation}
so that the order-by-order recurrence relation is
\begin{equation} \label{eqn:scalar-bn-order-by-order}
n(n-2l-1)b_n^{(r)}=\sum_{m=0}^{n-1}\sum_{k=0}^{r}B^{(k)}_{n,m,l,N}b_m^{(r-k)} \,.
\end{equation}
Notice that, by construction, $B^{(r)}_{n,m,l,N}=0$ when $r> n-m-rN$, i.e.\ $n-m-r(N+1)<0$. Next we prove the following result.

\noindent\textbf{Theorem 2.} The $\mathcal{O}(\alpha^r)$ contribution to $b_n$ is absent when $n< r(N+1)$ and to $a_0$ when $2l+1< r(N+1)$.

\noindent\textit{Proof.} This is true for $r=1$ from Theorem 1. We proceed by induction, assuming the result holds for some $r-1$. In \eqref{eqn:scalar-bn-order-by-order}, the term $B^{(k)}_{n,m,l,N}=0$ when $n-m-k(N+1)<0$, and $b_m^{(r-k)}=0$ when $m<(r-k)(N+1)$ by the inductive assumption, except for $b_m^{(r)}$. Therefore the product is zero when $m> n-k(N+1)$ or $m<(r-k)(N+1)$, so when $(r-k)(N+1)> n-k(N+1)$, or equivalently $n< r(N+1)$, the sum is zero for any $m$. The $b_m^{(r)}$ terms do not affect this conclusion because of the boundary condition $b_0^{(r)}=0$, so indeed $b_n^{(r)}=0$ $\forall n\leq r(N+1)$ if the lower-order terms vanish when $n< r(N+1)$. This proves the theorem concerning $b_n$.

Since the expression for $a_0$ at $\mathcal{O}(\alpha^r)$ is
\begin{equation*}
a_0^{(r)}\propto\sum_{m=0}^{2l}\sum_{k=0}^{r}B^{(k)}_{2l+1,m,l,N}b_m^{(r-k)} \,,
\end{equation*}
an identical proof follows with $n=2l+1$, concluding the proof.

This theorem establishes a sufficient condition for $a_0$ to vanish. Based on extensive explicit checks, we conjecture that a necessary condition also exists, although we have no rigorous proof of this.\footnote{Specifically, we conjecture that the bound $2l+1<r(N+1)$ is a necessary and sufficient condition for $a_0(l)$ to vanish at $\mathcal{O}(\alpha^r)$ if $r$ is odd. If $r$ is even, then $2l+1<r(N+1)$ is only a sufficient condition, while the necessary condition is given by the weaker bound $2l+3<r(N+1)$ in this case.}

A corollary of Theorem 2 is that, for any given $l$, only a finite number of terms can appear, i.e.\ the logarithmic Love number $a_0(l)$ is a polynomial in $\alpha$ for the metric \eqref{eq:single term metric fg}. In particular, for any $N\geq 2$, there exists $l$ such that $a_0(l)=0\;\Rightarrow\;\alpha=0$ ($N=1$ is an exception, covered by the known result for the RN black hole). In other words, while a cancellation may occur among different orders in $\alpha$ for some $l$, this can never happen for all $l$. This leads to the conclusion that any metric of the form \eqref{eq:single term metric fg} cannot have vanishing logarithmic Love number for all multipoles $l$.

\noindent\textit{Alternative expression for $a_0$.---} As a brief aside, we mention that the explicit result \eqref{eq:final a0 order alpha} for $a_0$ at $\mathcal{O}(\alpha)$ may be expressed as
\beq\bal
a_0^{(1)}&=\lim_{\epsilon\to0}\frac{N!\Gamma(N-2\hat{l})\Gamma^2(\hat{l}+1)\hat{l}}{\Gamma(2\hat{l}+2)\Gamma(2\hat{l}+1)\Gamma(-2\hat{l})\Gamma(N-\hat{l}+1)\Gamma(N-\hat{l})}\\
&\quad \times{}_4F_3(-\hat{l},-\hat{l}+1,N-2\hat{l},N+1;N-\hat{l}+1,N-\hat{l},-2\hat{l};1) \,,
\eal\eeq
where $\hat{l}:=(1+\epsilon)l$ is introduced to eliminate singularities within the expression.\footnote{Equivalently, we may express $a_0^{(1)}$ in terms of the so-called `regularized' hypergeometric function, as defined for instance in \href{https://mathworld.wolfram.com/RegularizedHypergeometricFunction.html}{Mathematica}.}

\subsubsection{Single-term deformation with $f\neq g$}

The assumption $f=g$ is inessential for our main result concerning the vanishing of $a_0$ at leading order in the metric deformation. If we consider metrics with\footnote{We assume there is no large hierarchy between $\alpha$ and $\beta$, so that either $|\alpha|x^N\gg |\beta|x^M$ or $|\alpha|x^N\ll |\beta|x^M$ (for $N\neq M$) everywhere in the domain of interest, i.e.\ sufficiently far from the black hole event horizon. This is of course consistent with the assumption that the horizon, on which $f=g=0$, must be perturbatively close to $x=1$.}
\beq \label{eq:two term metric fg}
f=1-x\left(1+\alpha x^N\right) \,,\qquad g=1-x\left(1+\beta x^M\right) \,,
\eeq
then a generalization of Theorem 2 holds.

We find that the coefficient of $\alpha^r\beta^s$ in the sum of \eqref{eq:b recurrence} is modified to
\begin{equation}\label{eqn:f-neq-g-B}\bal
B^{(r,s)}_{n,m,l,N,M}&:=\frac12\frac{(m-l)(n-m)}{n-m-rN}\bin{n-m-rN}{r}+\frac12\frac{(m-l)(n-m)}{n-m-sM}\bin{n-m-sM}{s} \\
&\quad +l(l+1)\bin{n-m-sM}{s} \,,
\eal\end{equation}
in the case $f\neq g$. We may now easily repeat the reasoning leading to Theorem 2. We notice that $B^{(r,s)}_{n,m,l,N,M}$ vanishes when both $n-m-r(N+1)<0$ and $n-m-s(M+1)<0$. Hence, depending on whether $r(N+1)$ is greater or less than $s(M+1)$, the proof of Theorem 2 can be replicated, with the conclusion that $a_0=0$ at $\mathcal{O}(\alpha^r\beta^s)$ when $2l+1<\min\{r(N+1),s(M+1)\}$.

\subsubsection{General deformation with $f=g$}

Given a metric with arbitrary functions $f$ and $g$, which we only assume to be analytic, i.e.\
\beq \label{eq:general series metric fg}
f=1-x\left(1+\sum_{N=1}^{\infty}\alpha_N x^N\right) \,,\qquad g=1-x\left(1+\sum_{N=1}^{\infty}\beta_N x^N\right) \,,
\eeq
we have not been able to find closed-form expressions for the corresponding coefficients $p_n$ and $q_n$. Nevertheless, we can still draw some general conclusions on the logarithmic Love number. For simplicity, we will set $g=f$ in what follows, although this assumption is inessential.

Consider first, for the sake of concreteness, a two-term deformation given by $f=1-x(1+\alpha x^N+\beta x^M)$, where we make no assumption on the sizes of $\alpha$ and $\beta$. Recall that $p$ and $q$ are given by \eqref{eq:pq when g equals f} when $g=f$. Thus it suffices to calculate
\beq
x\dfrac{d}{dx}\log{f}=-\sum_{n=1}^\infty\sum_{r=0}^{\lfloor{\frac{n}{N}}\rfloor}\sum_{s=0}^{\lfloor{\frac{n-rN}{M}}\rfloor}\frac{n}{n-rN-sM}\bin{n-rN-sM}{r}\bin{n-r(N+1)-sM}{s}\alpha^r\beta^s x^n \,,
\eeq
and
\beq
\frac{1}{f}=\sum_{n=1}^\infty\sum_{r=0}^{\lfloor{\frac{n}{N}}\rfloor}\sum_{s=0}^{\lfloor{\frac{n-rN}{M}}\rfloor}\bin{n-rN-sM}{r}\bin{n-r(N+1)-sM}{s}\alpha^r\beta^s x^n \,,
\eeq
from which it follows that $p_n^{(r,s)}$ and $q_n^{(r,s)}$ (the $\mathcal{O}(\alpha^r\beta^s)$ coefficients of $p_n$ and $q_n$, respectively) are both zero when $n<r(N+1)+s(M+1)$. Using a similar reasoning as in the previous subsections, we infer that $b_n^{(r,s)}=0$ for $n<r(N+1)+s(M+1)$ and hence that, in precise analogy with Theorem 2, $a_0=0$ at $\mathcal{O}(\alpha^r\beta^s)$ when $2l+1<r(N+1)+s(M+1)$.

The previous analysis may be easily generalized to arbitrarily, possibly infinitely, terms in the metric \eqref{eq:general series metric fg}. The above formulas indeed generalize in an obvious way, leading to the conclusion that $a_0=0$ at $\mathcal{O}(\prod_N\alpha_N^{r_N})$ if $2l+1<\sum_N r_N(N+1)$, for any given finite sequence $\{r_N\}$ of non-negative integers. We emphasize that these results demonstrate sufficient conditions for $a_0$ to vanish. Establishing necessary conditions appears much more challenging.

One strategy to partially address this question is to consider the solution of the equation $a_0(l)=0$ order by order in $l$. For any given level $l$, the expression is in general a non-linear polynomial of the coefficients $\alpha_N$, however by utilizing the constraints from the lower levels in the tower of equations we are able to simplify the constraint at level $l$. We remark that this argument is fully non-perturbative in the constants $\alpha_N$, a fact which will allow us to find counter-examples to the theorem on the vanishing of $a_0$.

Through direct calculation we can show the following proposition. Suppose $a_0(l')=0$ for all $l'\leq l$. Then
\beq
\sum_{m=l'+1}^{2l'}(-1)^m\bin{l'-1}{m-l'-1}\alpha_m=0 \,,\qquad 1\leq l'\leq l \,.
\eeq
In other words, the set of (non-linear) equations $\{a_0(l')=0\}_{l'=1}^{l-1}$ may be rewritten as a linear algebraic system for the variables $\{\alpha_2,\ldots ,\alpha_l\}$. This system is of course inhomogeneous as it also involves the variables $\{\alpha_{l+1},\ldots ,\alpha_{2l}\}$. If we were able to truncate the series, say at $\mathcal{O}(\alpha^L)$, then we arrive at a homogeneous system for the variables $\{\alpha_2,\ldots ,\alpha_L\}$. From the previous equation we see that the matrix for this system is upper triangular, hence non-singular. Thus $\alpha_2=\alpha_3=\cdots=\alpha_L=0$ is the unique solution.

It is tempting to think that this argument should apply to the infinite case. This is not so, and in fact it is not hard to see that the solution space of the whole set of constraints $\{a_0(l)=0\}_{l=1}^{\infty}$ is infinitely large. One obvious non-trivial solution is given by $\alpha_2=0,\alpha_3=\alpha_4=\cdots\equiv \alpha$, producing
\begin{equation}
f=1-x-\alpha_1x^2-\alpha\frac{x^4}{1-x} \,.
\end{equation}
This metric is inequivalent to Schwarzschild (and RN) and yet has exactly zero scalar logarithmic Love number for all multipoles $l$. Consistent with our previous theorems, this metric cannot be regarded as a perturbative deformation of Schwarzschild. This is clear from the fact that $f$ diverges at $x=1$, so that the event horizon (if any) cannot be perturbatively close to the Schwarzschild radius. In fact, this metric has a curvature singularity at $x=1$ (or $r=r_g$). For $\alpha>0$, there is a single Killing horizon in the domain $r>r_g$, located at $r_h=\frac{r_g}{2}\left(1+\sqrt{1+4\sqrt{\alpha}}\right)$ (setting the uninteresting term $\alpha_1=0$ for simplicity). For $\alpha<0$ the metric describes a naked singularity.

It would be interesting to see if there is any physical (or even mathematical) significance to this class of metrics characterized by zero running of the Love number. Naturally, of course, we do not expect the Love number to vanish beyond the logarithmic part. Furthermore, a vanishing running of the scalar Love number does not imply, of course, that the same holds for other types of perturbations; see our comments in Sec.\ \ref{sec:conclusion} in relation to these questions.

\section{Tidal response}

In this section we study the master formula \eqref{eq:a0 formula} for the logarithmic Love number in the case of odd-parity tensor perturbations in the scalar-tensor EFT, Eq.\ \eqref{eq:modified RW}. As we are mainly interested in a proof of principle of our method, we will keep the analysis as simple as possible in this section. We will therefore focus on leading order deformations of the Schwarzschild spacetime and also consider a concrete example.

\subsection{Perturbative deformation of the Schwarzschild metric} \label{subsec:tensor perturbation analysis}

We consider metrics with $g=f$ assumed to be perturbatively close to the Schwarzschild line element, and we will work at leading order in the deformation. One naturally expects the deformation in the function $\alpha_T$ to be also perturbative in this case, so we write
\beq \label{eq:tensor functions gamma theta}
f=1-x\left(1+\gamma x^N\right) \,,\qquad \alpha_T=\theta x^M \,,
\eeq
and we expect $\theta$ to be proportional to a positive (possibly non-integer) power of $\gamma$. For the time being, we make no assumption on the size of the ratio $\gamma/\theta$ and simply work to first order in both coefficients.

Substituting the above ansatz into \eqref{eq:modified RW} and expanding we read off the functions $p$ and $q$,
\beq\bal
p&=2-\frac{x}{1-x}+\gamma x^{N+1}\frac{(N x-N-1)}{(1-x)^2} -\theta x^M\frac{\left(M x^2-M-2 x\right)}{2 (1-x)^2} \,,\\
q&=-2-\frac{l(l+1)-2-x}{1-x}-\gamma x^{N+1}\frac{\left(l(l+1)+N
	x-N-3\right)}{(1-x)^2} \\
	&\quad -\theta x^M\frac{\left(-4 xl(l+1)+M^2 x^2-2 M^2 x+M^2+2 M x^2-7 M x+5 M+12 x\right)}{4 (1-x)^2} \,.
\eal\eeq
It is now straightforward, if more tedious, to repeat the same steps of Sec.\ \ref{subsec:scalar deformation}. We then eventually arrive at a closed-form expression for the logarithmic Love number $a_0$, with terms proportional to $\gamma$ and $\theta$ at the order we are working. Writing $a_0=\gamma a_0^{(\gamma)}+\theta a_0^{(\theta)}+\ldots\,$, we have explicitly
\beq
\label{eqn:odd-parity-a0-gamma}
a_0^{(\gamma)} =-\frac{l+2}{2l+1}\sum_{m=N}^{2l}\frac{\pochh{m-l+2}{2l-m}\pochh{m-l-1}{2l-m}\pochh{-l+2}{m-N}\pochh{-l-2}{m-N+1}}{\pochh{m+1}{2l-m}\pochh{m-2l}{2l-m}(m-N)!\pochh{-2l}{m-N}} \,,
\eeq
\beq\bal \label{eqn:odd-parity-a0-theta}
a_0^{(\theta)}&=\frac{1}{2l+1}\sum_{m=M}^{2l}\left[(m-l)\left(m-l-\frac32M\right)+\frac34M^2-4\right] \\
&\quad\times \frac{\pochh{-l+2}{m-M}\pochh{-l-2}{m-M}\pochh{m-l+3}{2l-m}\pochh{m-l-1}{2l-m}}{(m-M)!\pochh{-2l}{m-M}\pochh{m+1}{2l-m}\pochh{m-2l}{2l-m}} \\
&\quad+\frac{1}{2l+1}\sum_{m=M-1}^{2l}\frac{M}{2}\left[m-l-\frac32M-\frac52\right] \\
&\quad\times \frac{\pochh{-l+2}{m-M+1}\pochh{-l-2}{m-M+1}\pochh{m-l+3}{2l-m}\pochh{m-l-1}{2l-m}}{(m-M+1)!\pochh{-2l}{m-M+1}\pochh{m+1}{2l-m}\pochh{m-2l}{2l-m}} \,.
\eal\eeq

Notice that all these results assume $l\geq 2$, since indeed the modified Regge-Wheeler equation \eqref{eq:modified RW} is only applicable in this case. Tidal Love numbers with $l=0$ and $l=1$ are not defined, since the corresponding tensor modes are pure gauge once the global charges of the spacetime are fixed. This can also be understood mathematically from the fact that tensor spherical harmonics only exist for $l\geq 2$.

\subsubsection{$\gamma\gg \theta$}

In this case the leading order term is given by $a_0^{(\gamma)}$. The key observation is that each summand in \eqref{eqn:odd-parity-a0-gamma} is sign-definite, hence there can be no cancellation among individual terms.

To see this, it suffices to examine the sign of each term in \eqref{eqn:odd-parity-a0-gamma} (ignoring $\pochh{m+1}{2l-m}$ and $(m-N)!$, which are trivially positive):
\begin{equation*}
\sgn \pochh{m-l+2}{2l-m}=\begin{cases}
    0 & \mbox{if $m-l+2\leq0$} \,, \\
    1 & \mbox{if $m-l+2>0$} \,, \\
\end{cases}
\qquad\sgn \pochh{m-l-1}{2l-m}=\begin{cases}
    0 & \mbox{if $m-l-1\leq0$} \,, \\
    1 & \mbox{if $m-l-1>0$} \,, \\
\end{cases}
\end{equation*}
\begin{equation*}
\sgn \pochh{-l+2}{m-N}=\begin{cases}
    0 & \mbox{if $m-l-N+1\geq0$} \,, \\
    (-1)^{m-N} & \mbox{if $m-l-N+1<0$} \,, \\
\end{cases}
\end{equation*}
\begin{equation*}
\sgn \pochh{-l-2}{m-N+1}=\begin{cases}
    0 & \mbox{if $m-l-N-2\geq0$} \,, \\
    (-1)^{m-N+1} & \mbox{if $m-l-N-2<0$} \,, \\
\end{cases}
\end{equation*}
\begin{equation*}
\sgn \pochh{m-2l}{2l-m}=(-1)^m \,,\qquad\sgn \pochh{-2l}{m-N}=(-1)^{m-N} \,.
\end{equation*}
Collecting these, we infer that the $m$-th summand in \eqref{eqn:odd-parity-a0-gamma} is non-zero for $m\in[\![l+2,l+N-2]\!]\cap[\![N,2l]\!]$, with $m$-independent sign $(-1)^N$. Notice that $2l\geq l+2$ and $N\leq l+N-2$ when $l\geq2$. Therefore, demanding $[\![l+2,l+N-2]\!]\cap[\![N,2l]\!]\neq\varnothing$, combined with $l\geq2$, yields $N\geq4$ and $l\geq \left\lceil\frac{N}{2}\right\rceil$ as necessary and sufficient conditions to have $a_0^{(\gamma)}(l)\neq0$ at $\mathcal{O}(\gamma)$.

\subsubsection{$\gamma\ll \theta$}

In this case the leading order term is given by $a_0^{(\theta)}$. The analysis is very analogous, only more tedious due to the presence of two sums in \eqref{eqn:odd-parity-a0-theta}. As before, the outcome depends on $M$, and it proves convenient to analyze the different cases separately.

Suppose $M\leq3$. Then the product $\sgn \pochh{m-l-1}{2l-m}\pochh{2-l}{m-M}$ in the first sum is zero for each summand. If $M=4$, then this product is non-zero only when $m=l+2$, but in this case the expression in square brackets happens to vanish, so every summand is still zero. A very similar reasoning shows that the second sum also vanishes when $M \leq 4$. 

Suppose next $M>4$. In the first sum, the term in brackets can be shown to be strictly positive, while the remaining part of the $m$-th summand is non-zero only when $m\in[\![l+2,M+l-2]\!]$, in which case it has sign $(-1)^M$, independent of $m$. In the second sum, one can similarly check that the $m$-th summand is non-zero only when $m\in[\![l+2,M+l-3]\!]$, in which case its sign is $(-1)^M$ ($(-1)$ from the term in brackets and $(-1)^{M+1}$ from the ratio of factorials). There can therefore be no cancellation among individual terms in this case.

In conclusion, it is necessary and sufficient that $M\geq 5$ and $l\geq \left\lceil\frac{M-1}{2}\right\rceil$ to have to have $a_0^{(\theta)}(l)\neq0$ at $\mathcal{O}(\theta)$.

\subsubsection{$\gamma\sim \theta$} \label{subsubsec:gamma sim theta}

In this case one cannot rule out a cancellation between $a_0^{(\gamma)}$ and $a_0^{(\theta)}$. However, it should be clear that such a cancellation cannot occur for all $l$, so we can still conclude that logarithmic running must show up at some multipole order. The explicit proof of this statement is straightforward but tedious, so we relegate it to the appendix for the interested reader. 

Notice that the trivial-$a_0$ thresholds established above still of course hold in this case. That is, $a_0$ necessarily vanishes at leading order if $N<4$ and $M<5$. This is consistent with the findings of \cite{Wang:2025oek} based on an order-by-order calculation of the Love numbers for some explicit regular black hole metrics. On the other hand, our results show that logarithmic running can in fact appear at leading order.

\subsection{Illustration: Hayward black hole}

The usefulness of our result may be illustrated with a concrete example. We consider here the Hayward regular black hole metric~\cite{Hayward:2005gi}, whose Love numbers were calculated order by order in the deformation parameter in Ref.~\cite{Barura:2024uog} in the context of the EFT of perturbations in scalar-tensor theories. Our method allows to derive exact, closed-form results for the logarithmic part of the Love number.

The Hayward metric functions are given by
\beq\bal \label{eq:hayward functions}
&f=1-\frac{x}{(1+\eta)(\eta x^3+1)}\,,\qquad \alpha_T=-\frac{\eta x^3(2+\eta x^3)}{(\eta x^3+1)^2}\,,\\
&F=(1-x)\frac{\eta x^3+\eta x^2+\eta x+\eta+1}{(1+\eta)(\eta x^3+1)} \,,
\eal\eeq
where $\eta>0$ is the parameter controlling the deviation from GR. This system falls into the case of Sec.\ \ref{subsubsec:gamma sim theta} with $N=M=3$.\footnote{This is not exactly correct because the parameterization of \eqref{eq:hayward functions} implies a deformation at $\mathcal{O}(x^0)$. Since our calculations here are non-perturbative in the metric deformation, this issue is immaterial.} Inserting \eqref{eq:hayward functions} into \eqref{eq:modified RW} we find
\beq\bal
p&=-\frac{3 \eta ^2 x^7+6 \eta  x^4+\eta  (\eta +1) x^3+3 x-2 (\eta +1)}{(1-x) \left(\eta  x^3+1\right) \left(\eta  (x+1) \left(x^2+1\right)+1\right)} \,,\\
q&=-\frac{(\eta +1) \left(\eta  x^3+1\right) \left(\frac{4 \left(l^2+l-2\right)}{\left(\eta  x^3+1\right)}+\frac{3 x \left(4 \eta  x^3-5\right)}{(\eta +1) \left(\eta 
   x^3+1\right)^2}+\frac{\eta  x^3 \left(5 \eta  x^3-32\right)+8}{\left(\eta  x^3+1\right)^3}+\frac{3 x}{\eta +1}\right)}{4 \left(\eta -\eta  x^4+1-x\right)} \,.
\eal\eeq

Although we have not attempted to obtain closed-form results for general $l$, for any given $l$ the calculation of $a_0$ is completely mechanical: (i) expand $p$ and $q$ in Taylor series up to order $\mathcal{O}(x^{2l+1})$; (ii) compute $\{b_1,b_2,\ldots,b_{2l}\}$ recursively from \eqref{eq:b recurrence}; (iii) substitute in \eqref{eq:a0 formula}.

For $l=2,3,4,5$ we obtain
\beq
a_0(2)=0 \,,\qquad a_0(3)=0 \,,\qquad a_0(4)=-\frac{24 \eta ^2}{25 (\eta +1)^3} \,,\qquad a_0(5)=-\frac{6909 \eta ^2}{6875 (\eta +1)^5} \,.
\eeq
The vanishing of $a_0$ for $l=2,3$ agrees with the calculations of~\cite{Barura:2024uog} and with the results of Sec.\ \ref{subsubsec:gamma sim theta}, which established that $a_0$ must vanish at $\mathcal{O}(\eta)$. We emphasize however that our results here are valid to all orders in $\eta$. Our expression for $a_0(4)$ is also in exact agreement with~\cite{Barura:2024uog}, which found $a_0(4)=-\frac{24}{25}\eta^2+\mathcal{O}(\eta^3)$ in a perturbative calculation. The result for $a_0(5)$ is new to the best of our knowledge. The fact that we can derive non-perturbative results with little effort underscores the advantage of our method, insofar of course one is only interested in the logarithmic running.


\section{Discussion} \label{sec:conclusion}

Our aim with this paper was to highlight and exploit the fact that the logarithmic running of the static linear response coefficients of black holes can be calculated directly from Fuchsian theory, without knowledge of the general solution or the need of connection formulas for special functions. The calculation of the logarithmic Love number via Eq.\ \eqref{eq:a0 formula} is completely straightforward and mechanical, applicable for any given spacetime. We illustrated this fact by recovering the known results for the Schwarzschild-Tangherlini and Hayward black holes. Furthermore, our results highlight a perhaps surprising fact: the logarithmic running is agnostic about boundary conditions at the event horizon. This appears to be consistent with the interpretation of this term as a beta function, i.e.\ a property of the theory rather than of the detailed structure of the black hole's horizons and singularities. We believe this point deserves further investigation.

We see no fundamental obstruction to the generalization of our methods to more complex systems. In future work we plan to analyze the case of coupled equations, which is relevant for metrics sourced by matter fields. The setting of spinning black holes is also within the immediate scope of our techniques provided the equations are separable. It is perhaps worth emphasizing that all our results are only valid in linear response theory. However, non-linear Love numbers~\cite{Riva:2023rcm,Iteanu:2024dvx,DeLuca:2023mio,Kehagias:2024rtz,Gounis:2024hcm,Pani:2025qxs} could in principle be tackled in perturbation theory, which is anyway the more proper setting to discuss the effects of logarithmic running. This approach would hopefully complement parallel efforts in the context of the world-line EFT of black holes. On the other hand, one aspect which is \textit{not} immediately encompassed by our methods is the linear response of nonzero frequency perturbations, i.e.\ dynamical Love numbers, since in this case the boundary $x=0$ is not a regular singular point; rather, it is irregular, and the Frobenius series method is not directly applicable. We hope to address these questions in future work.

One motivation for the present work was the question of establishing necessary and sufficient conditions for the logarithmic Love number to vanish. This problem was recently addressed in~\cite{Sharma:2025xii} which concluded that any perturbative deformation of the Schwarzschild or RN metrics must have non-zero Love numbers. Our study, which fully corroborates this conclusion, underscores that the perturbativity assumption is critical: at least for the logarithmic term, metrics do exist that have exactly zero Love number for all multipoles. Once again, the same outcome is not expected to uphold for the non-running Love numbers, and indeed Ref.~\cite{Sharma:2025xii} has argued, based on the uniqueness of ladder symmetries for the Schwarzschild and RN spacetimes~\cite{Sharma:2024hlz}, that any modification of these metrics must have non-zero Love numbers. We might comment here that this is also related to the link between ladder symmetries and the `dual' formulation, valid in the static limit, of the Klein-Gordon equation on the background metric~\cite{Hui:2022vbh}
\beq
\d\tilde{s}^2=-\frac{\Delta}{L^2}\d \tilde{t}^{\,2}+\frac{L^2}{\Delta}\d \tilde{r}^2+L^2\d\Omega^2_{S^2}\,,
\eeq
where $\Delta:= \tilde{r}^2f(\tilde{r})$ in terms of the function $f$ that defines the original metric \eqref{eq:general metric fg}. Here we set $g=f$ for simplicity, and notice that $L$ may be more generally taken to be a function of the radial coordinate. The ladder symmetries appear to be intimately related to the isometries of the $(\tilde{t},\tilde{r})$ sub-manifold~\cite{Combaluzier-Szteinsznaider:2024sgb}: the requirement that this 2-dimensional space must be anti-de Sitter selects the RN (or Schwarzschild) metric function $f$ as the unique solution. In this vein, it would be interesting to explore if spacetimes with zero logarithmic running may be understood from some (possibly reduced) symmetry considerations.

\subsubsection*{Acknowledgments}

We are very grateful to Ao Guo and Xinmiao Wang for collaboration in the early stages of this project, and to Jun Wang and Sam Wong for useful conversations and comments on the draft. SGS acknowledges support from the NSFC (Grant No.\ 12250410250) and from a Provincial Grant (Grant No.\ 2023QN10X389).

\appendix

\section{Analysis of the case $\gamma\sim\theta$}

In this appendix we provide the analysis of the case where $\gamma$ and $\theta$ in Eq.\ \eqref{eq:tensor functions gamma theta} are of the same order. We wish to prove that the logarithmic Love number cannot vanish at leading order in $\gamma\sim\theta$ for all multipoles $l$, except when $N<4$ and $M<5$. From the analysis of Sec.\ \ref{subsec:tensor perturbation analysis}, we know that both $a_0^{(\gamma)}$ and $a_0^{(\theta)}$, cf.\ \eqref{eqn:odd-parity-a0-gamma} and \eqref{eqn:odd-parity-a0-theta}, are non-zero in the non-trivial range $l\geq \max\left\{\left\lceil\frac{N}{2}\right\rceil,\left\lceil\frac{M-1}{2}\right\rceil\right\}$.

First, notice that if $N\geq M+1$, then the choice $l=\left\lceil\frac{N}{2}\right\rceil$ yields $a_0^{(\gamma)}\neq0$ and $a_0^{(\theta)}=0$. Thus a cancellation cannot occur in this case. Similarly, if $M\geq N+3$, then the choice $l=\left\lceil\frac{M-1}{2}\right\rceil$ yields $a_0^{(\gamma)}=0$ and $a_0^{(\theta)}\neq0$. We may therefore disregard these two cases. This leaves us with three possibilities: $M=N$, $M=N+1$ and $M=N+2$. If $M=N$ is an odd number, then $\left\lceil\frac{N}{2}\right\rceil>\left\lceil\frac{M-1}{2}\right\rceil$, so again we have $a_0^{(\gamma)}\neq0$ and $a_0^{(\theta)}=0$ for $l=\left\lceil\frac{N}{2}\right\rceil$. Similarly, if $M=N+2$ is an even number, the same conclusion follows. The case $M=N+1$ is not restricted in this way.

All in all we have four non-trivial cases: (i) $M=N$, an even number; (ii) $M=N+1$, with odd $N$; (iii) $M=N+1$, with even $N$; (iv) $M=N+2$, an odd number. In these cases $a_0^{(\gamma)}(l)\neq0$ if and only if $a_0^{(\theta)}(l)\neq0$.

We now argue by contradiction, i.e.\ assume that a cancellation occurs for all $l$ in the non-trivial range. If this is true, then we must have
\beq \label{eq:app ratio a0theta-a0gamma}
\frac{a_0^{(\theta)}(l)}{a_0^{(\gamma)}(l)}=\frac{a_0^{(\theta)}(l')}{a_0^{(\gamma)}(l')} \,,
\eeq
for all $l,l'$ in this range. In particular, it must be true for $l=\max\left\{\left\lceil\frac{N}{2}\right\rceil,\left\lceil\frac{M-1}{2}\right\rceil\right\}=\left\lceil\frac{N}{2}\right\rceil$ and $l'=l+1$. It is then easy to infer a contradiction for each of the above four cases:

\begin{itemize}
    \item[] Case (i): With $N=M=2l$, $l\geq3$, Eq.\ \eqref{eq:app ratio a0theta-a0gamma} yields $(l+1)(2l-3)=0$, a contradiction;

    \item[] Case (ii): With $N=M-1=2l-1$, $l\geq3$, Eq.\ \eqref{eq:app ratio a0theta-a0gamma} yields $(l+1)(2l-3)=0$, a contradiction;

    \item[] Case (iii): With $N=M-1=2l$, $l\geq2$, Eq.\ \eqref{eq:app ratio a0theta-a0gamma} yields $(l+1)(l+2)(2l-3)=0$, a contradiction;

    \item[] Case (iv): With $N=M-2=2l-1$, $l\geq3$, Eq.\ \eqref{eq:app ratio a0theta-a0gamma} yields $(l-2)(l+1)(l+2)(l+3)(2l-3)=0$, a contradiction.
\end{itemize}
This concludes the proof.


\bibliographystyle{unsrturl}
\bibliography{LogLove_refs}

\end{document}